\documentclass{aa}
\usepackage[varg]{txfonts}
\usepackage{graphicx}
\usepackage{txfonts}
\usepackage{natbib}
\usepackage{longtable}
\usepackage{lscape}
\usepackage{multirow}
\bibpunct{(}{)}{;}{a}{}{,}

\newcommand{\kms}{km\,s$^{-1}$}

\newcommand{\mum}{$\mu$m}

\newcommand{\Msun}{$M_{\odot}$}

\newcommand{\accunit}{$M_{\odot}$\,yr$^{-1}$}
\newcommand{\rev}{ }
\newcommand{\newrev}{ }

\newcommand{\sect}{Sect.}

\newcommand{\Mdotacc}{$\dot M_{\rm{acc}}$}
\newcommand{\mjybeam}{mJy~beam$^{-1}$}

\begin{document}

\title{Millimeter Observations of the disk around GW~Ori}

\author{M.~Fang\inst{1, 2} \and A.~Sicilia-Aguilar\inst{3, 1} \and D.~Wilner\inst{4} \and Y.~Wang\inst{5, 6} \and  V.~Roccatagliata\inst{7} \and D.~Fedele\inst{8} \and J.~Z.~Wang\inst{9}}
\institute{Departamento de F\'{\i}sica Te\'{o}rica, Universidad Aut\'{o}noma de Madrid, Cantoblanco 28049, Madrid, Spain \and Department of Astronomy, University of Arizona, 933 North Cherry Avenue, Tucson, AZ 85721, USA
\and SUPA, School of Physics and Astronomy, University of St Andrews, North Haugh, St Andrews KY16 9SS, UK \and Harvard–Smithsonian Center for Astrophysics, 60 Garden Street, Cambridge, MA 02138, USA \and Max Planck Institute for Astronomy, K\"{o}nigstuhl 17, 69117, Heidelberg, Germany \and Purple Mountain Observatory and Key Laboratory of Radio Astronomy, Chinese Academy of Sciences, 2 West Beijing Road, 210008 Nanjing, China \and  Universit\"ats-Sternwarte M\"unchen, Ludwig-Maximilians-Universit\"at, Scheinerstr. 1, 81679, M\"unchen, Germany \and  Max Planck Institut f\"ur Extraterrestrische Physik, Giessenbachstrasse 1, 85748, Garching, Germany \and Shanghai Astronomical Observatory, Chinese Academy of Sciences, 80 Nandan Road, 200030, Shanghai, China}

\date{Received; Accepted}

\abstract{The GW~Ori system is a pre-main sequence triple system (GW~Ori~A/B/C) with  companions (GW~Ori~B/C) at $\sim$1\,AU and $\sim$8\,AU, respectively, from the primary (GW~Ori~A). The primary of the system has a mass of 3.9\,\Msun, but shows a spectral type of G8. Thus, GW~Ori~A could be a precursor of a B star, but it is still at an earlier evolutionary stage than Herbig~Be stars. GW~Ori provides us an ideal target for experiments and observations (being a ``blown-up'' upscaled Solar System with a very massive ``sun'' and at least two ``upscaled planets''). We present the first spatially-resolved millimeter interferometric observations of the disk around the triple pre-main-sequence system GW~Ori, obtained with the the Submillimeter Array, both in continuum and in the $^{12}{\rm CO}~J=2-1$, $^{13}{\rm CO}~J=2-1$, and ${\rm C^{18}O}~J=2-1$ lines. These new data reveal a huge, massive, and  bright disk in the GW~Ori system. The dust continuum emission suggests a disk radius around 400\,AU. But, the $^{12}{\rm CO}~J=2-1$ emission shows much more extended disk with a size around 1300\,AU. Due to the spatial resolution ($\sim$1$''$), we cannot detect the gap in the disk which is inferred from spectral energy distribution (SED) modeling. We characterize the dust and gas properties in the disk by comparing the observations with the predictions from the disk models with various parameters calculated with a Monte Carlo radiative transfer code  RADMC--3D. The disk mass is around {\rev 0.12}\,\Msun, and the disk inclination with respect to the line of sight is around $\sim${\rev 35}$^\circ$. The kinematics in the disk traced by the CO line emission strongly suggest that the circumstellar material in the disk is in Keplerian rotation around GW~Ori. {\rev Tentatively substantial ${\rm C^{18}O}$ depletion in gas phase is required to explain the characteristics of the line emission from the disk.}
}


\maketitle 

\section{Introduction}
Spectroscopic and imaging  surveys suggest that a majority of young stars are formed in  binary/multiple systems \citep{1993AJ....106.2005G,1993A&A...278..129L,1997ApJ...481..378G,2008ApJ...683..844L,2011ApJ...731....8K}. Theoretical and observational studies indicate  that the interaction between disks and companions is an efficient mechanism to dissipate disks \citep{1993prpl.conf..749L,2009ApJ...696L..84C,2012ApJ...745...19K}. Thus, it is very important to investigate disks around the binary or multiple stellar systems, in order to understand disk evolution, as well as planet formation. The triple system GW~Ori is an ideal target for such study.

GW~Ori is located at $\lambda$~Ori \citep[$\sim$400\,pc,][]{2013MNRAS.434..806B}, and was revealed as a triple stellar system recently \citep[GW~Ori~A/B/C,][]{1991AJ....101.2184M,2011A&A...529L...1B}. The primary GW~Ori~A is a G8 pre-main sequence star with a mass of $\sim$4\,\Msun, which makes it a very interesting system between Herbig~Be stars and classical T~Tauri stars \citep{2014A&A...570A.118F}. The close companion GW~Ori~B was discovered as a spectroscopic binary with an orbital period of $\sim$242\,days and a separation of $\sim$1\,AU \citep{1991AJ....101.2184M,2014A&A...570A.118F}. The second companion GW~Ori~C, located at a projected separation of $\sim$8\,AU from GW~Ori~A, was detected with near-infrared interferometric technique \citep{2011A&A...529L...1B}. The sub-millimeter and  millimeter observations show that GW~Ori is still harboring a massive disk \citep{1995AJ....109.2655M}, which is one of the most massive disks around a G-type star. Strong ongoing accretion activity (\Mdotacc$\sim$3--4$\times$10$^{-7}$\,\accunit ) from the disk to the central star(s) in the GW~Ori system was suggested from the $U$-band excess, and strong and broad H$\alpha$ and H$\beta$ emission lines on the spectrum of GW~Ori \citep{2004AJ....128.1294C,2014A&A...570A.118F}. 

In \citet[][hereafter Paper\,I]{2014A&A...570A.118F}, we presented a study of the inner disk in the GW~Ori system based on the infrared data. We reproduced the spectral energy distribution (SED) of GW~Ori using disk models with gaps sized 25--55\,AU. We found that the SED of GW~Ori exhibited dramatic changes on timescales of $\sim$20\,yr in the near-infrared bands, which can be interpreted as the change in the amount and distribution of dust particles in the gap due to a ``leaky dust filter''. Due to its brightness {\newrev at submillimeter and millimeter wavelengths}, the disk mass and size has been subject to lot of speculation with  sizes up to 500\,AU radius and  masses that could render it gravitational unstable \citep{1995AJ....109.2655M,2009A&A...502..367S}. In this work, we present an investigation of the outer disk around GW~Ori using the new millimeter data obtained from the Submillimeter Array \citep[SMA,][]{2004ApJ...616L...1H}. These new observations have spatially resolved the disk around GW~Ori for the first time. We arrange this paper as follows: In \sect~\ref{Sec:data} we describe the observations and data reduction. In \sect~\ref{Sec:obs_result} we present our observational results, and in  \sect~\ref{Sec:modeling} we describe the disk modeling, and compare the model results with the observations, which are then discussed in~\ref{Sec:discussion}. We summarize our results in \sect~\ref{Sec:summary}.

\begin{figure*}
\centering
\includegraphics[width=1.\columnwidth]{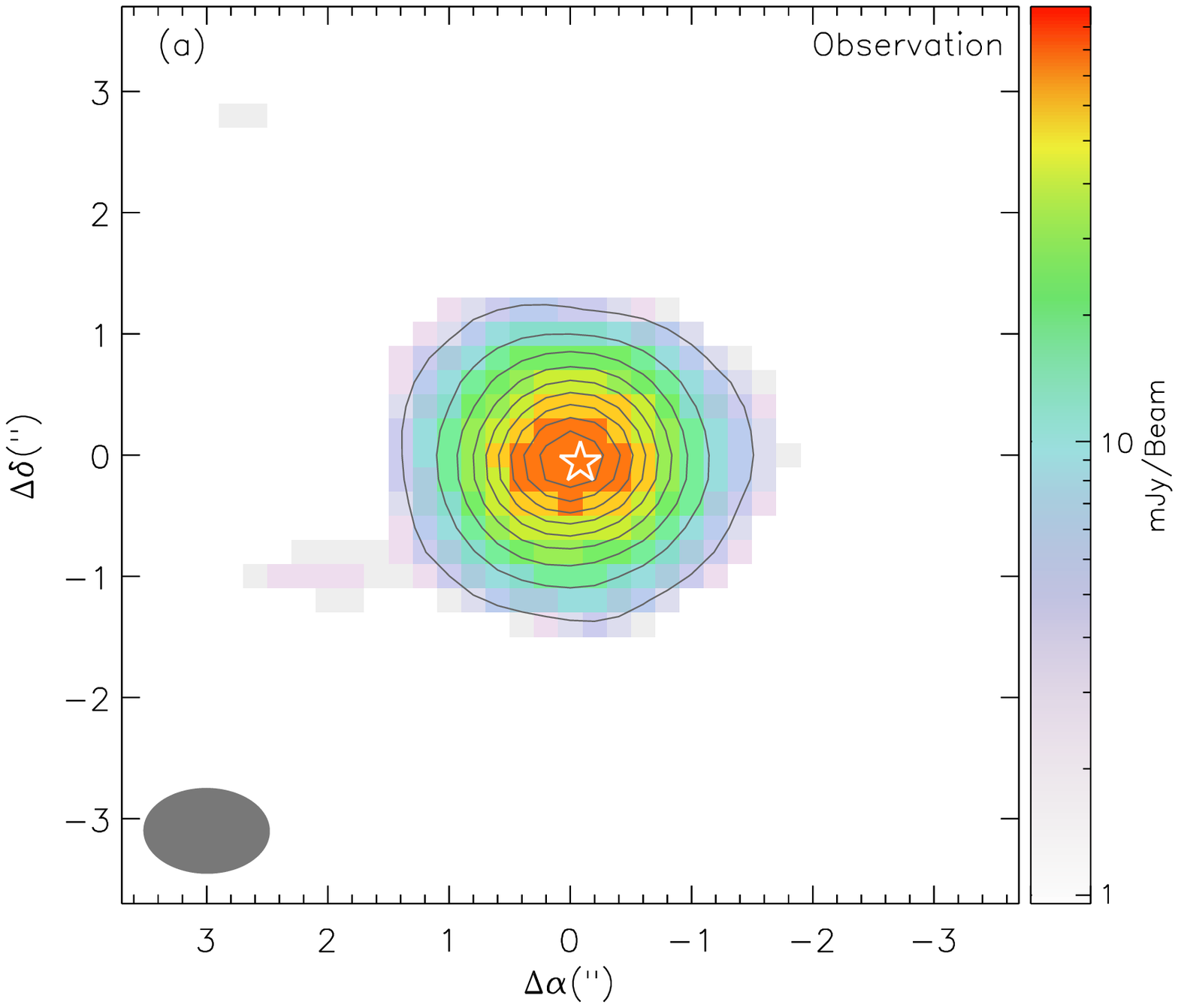}
\includegraphics[width=1.\columnwidth]{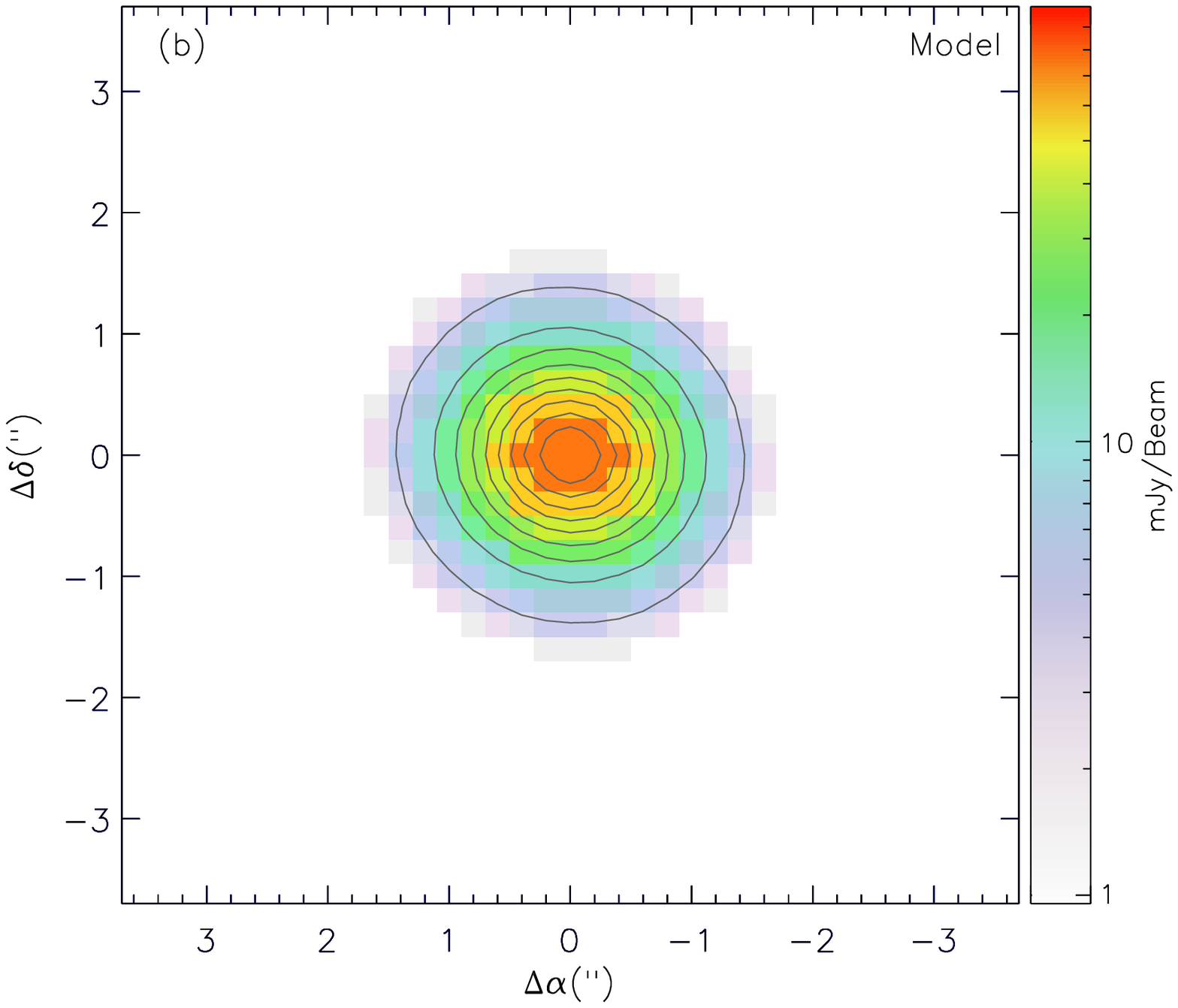}
\includegraphics[width=1\columnwidth]{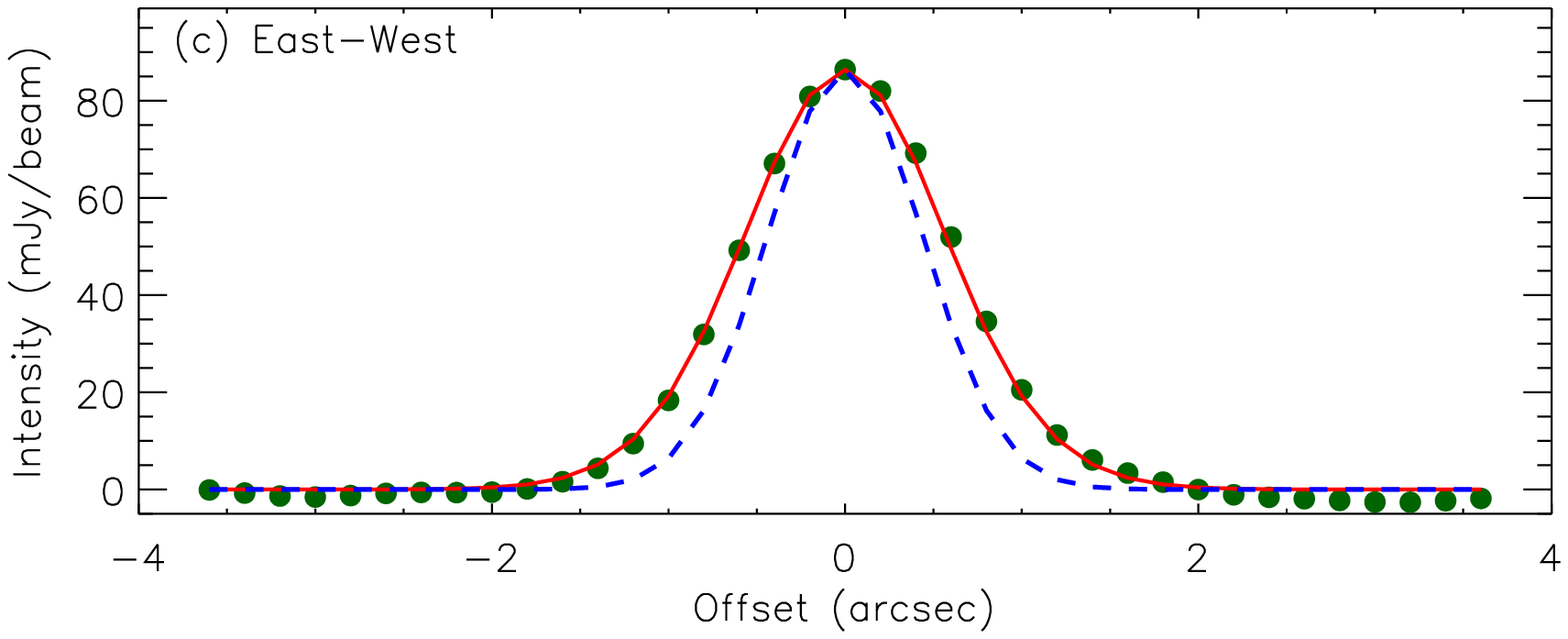}
\includegraphics[width=1\columnwidth]{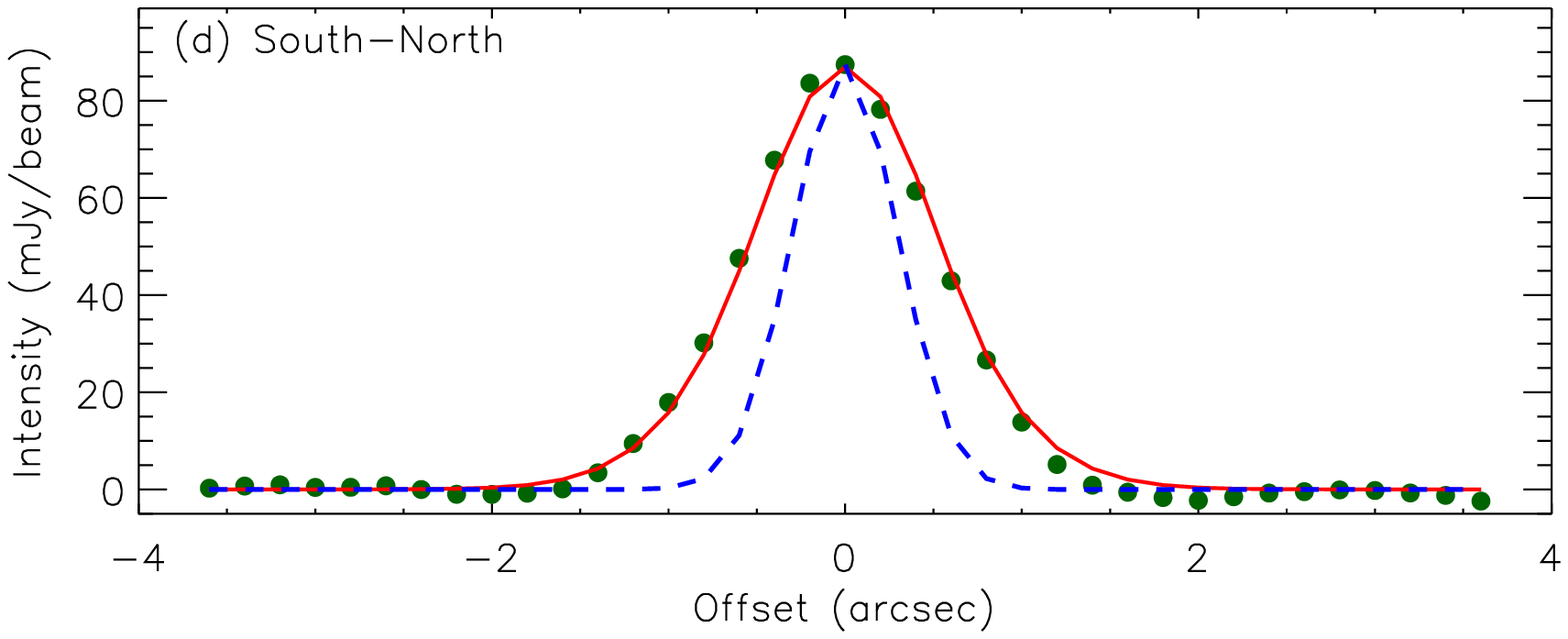}
\caption{(a) The observed  map of the continuum emission observed toward GW~Ori at the wavelength of 1.3~mm, with contours (solid lines) drawn at 9.5\,mJy~beam$^{-1}$ (10\,$\sigma$) intervals, starting at 4.8~mJy~beam$^{-1}$ (5\,$\sigma$).  The synthesized beam is shown in the lower left corner. The position of the central star is indicated with the star symbol. (b) The modeled map of the continuum emission for the GW~Ori disk system at  the wavelength of 1.3~mm. The contour levels are same as in Panel~(a). (c, d) The distribution of the observed intensities (filled circles) along the east-west (c) and  north-south (d) direction cross the center of the map compared with the our model (solid lines). The dash lines show the expected profiles for an unresolved object. \label{Fig:cont}}
\end{figure*}

\section{Observations, data reduction}\label{Sec:data}
The object GW~Ori was observed with the SMA on January 5, 2010 in the compact configuration with six antennas, on January 19, 2011 in the very extended configuration with six antennas, and  February 2, 2011 in the extended configuration with seven antennas. The phase center of the field was RA=05h29m08.38s and Dec=$+11^{\circ}52^{\prime}12.7^{\prime\prime}$ (J2000.0). The SMA has two spectral sidebands, both 4\,GHz wide and separated by 10\,GHz. The receivers were tuned to 230.538 GHz in the upper sideband ($v_{\rm lsr}$ = 11~km~s$^{-1}$) with a maximum spectral resolution of 1.1\,\kms\ on the upper sideband and 1.2~km~s$^{-1}$ on the lower sideband. The system temperatures ($T_{\rm sys}$) were around $\sim$80--200~K during the observation on January 5, 2010, $\sim$100--200~K on January 19, 2011, and $\sim$110--300~K on February 2, 2011. For the compact configuration on January 5, 2010, the bandpass was derived from the quasar 3c273 observations. Phase and amplitude were calibrated with regularly interleaved observations of the quasar  0530+135 (1.7$^{\circ}$ away from the source). The flux calibration was derived from Titan observations, and the flux scale is estimated to be accurate within 20\%. For the very extended configuration on January 19, 2011, the bandpass was derived from the quasar 3c279 observations. Phase and amplitude were calibrated with regularly interleaved observations of the quasar 0530+135 and 0423--013. The flux calibration was derived from Ganymede observations, and the flux scale is estimated to be accurate within 20\%. For the extended configuration on February 2, 2011, the bandpass was derived from the quasar 3c279 observations. Phase and amplitude were calibrated with regularly interleaved observations of the quasar 0530+135 and 0423--013. The flux calibration was derived from Titan observations, and the flux scale is estimated to be accurate within 20$\%$. 

We merged the three configuration data sets, applied different robust parameters for the continuum and line data, and got the synthesized beam sizes between 1.03$^{\prime\prime}$ $\times$0.70$^{\prime\prime}$ (PA$\sim$89.5$^\circ$) and 1.15$^{\prime\prime}\times0.83^{\prime\prime}$ (PA$\sim-$88.6$^\circ$), respectively. The rms of 1.3~mm continuum image is  0.96 mJy beam$^{-1}$, and the rms of the $^{12}$CO$~J=2-1$ data is 0.05~Jy~beam$^{-1}$ at 1.1\,\kms\ spectral resolution, 0.04~Jy~beam$^{-1}$ at 1.2\,\kms\ spectral resolution for $^{13}$CO~$J=2-1$, and 0.03~Jy~beam$^{-1}$ at 1.2\,\kms\ spectral resolution for C$^{18}$O~$J=2-1$. The flagging and calibration was done with the IDL superset MIR \citep{1993PASP..105.1482S}, which was originally developed for the Owens Valley Radio Observatory and adapted for the SMA\footnote{The MIR cookbook by Chunhua Qi can be found at \url{http://cfa-www.harvard.edu/~cqi/mircook.html}.}. The imaging and data analysis were conducted in MIRIAD \citep{1995ASPC...77..433S}.

\section{Observational Results}\label{Sec:obs_result}
\subsection{Dust continuum emission}
In Fig.~\ref{Fig:cont}(a), we show the continuum emission map of GW~Ori with contours starting at 4.8\,\mjybeam\ (5$\sigma$) and increasing at 9.5\mjybeam\ (10$\sigma$) intervals. Considering a 20\% systematic calibration uncertainty, the integrated continuum flux density in this map is 320$\pm$64\,mJy, consistent with the result (255$\pm$60\,mJy) in \citet{1995AJ....109.2655M}. From a two-dimensional Gaussian fit to the image, the full width at half maximum (FWHM) of the continuum emission is 1$\farcs$4($\pm0\farcs004$)$\times$1$\farcs$3($\pm0\farcs004$), suggesting that the continuum emission of GW~Ori is resolved given the synthesized beam size of 1.03$^{\prime\prime}$ $\times$0.70$^{\prime\prime}$. After the deconvolution of the synthesized beam, the FWHM of the  continuum emission map is 0$\farcs$9$\times$1$\farcs$1 which is corresponding to 360$\times$400\,AU at a distance 400\,pc. In Fig.~\ref{Fig:cont}(c, d) we show the distribution of the intensities for the the continuum emission map along the east-west and  north-south directions cross the center of the map, and the expected distrubutions for an unresolved object according to the spatial resolutions of our observations. We note that the emission of GW~Ori along the east-west direction is marginally resolved, but the one along the north-south direction is well resolved.
\begin{figure*}
\centering
\includegraphics[width=2\columnwidth]{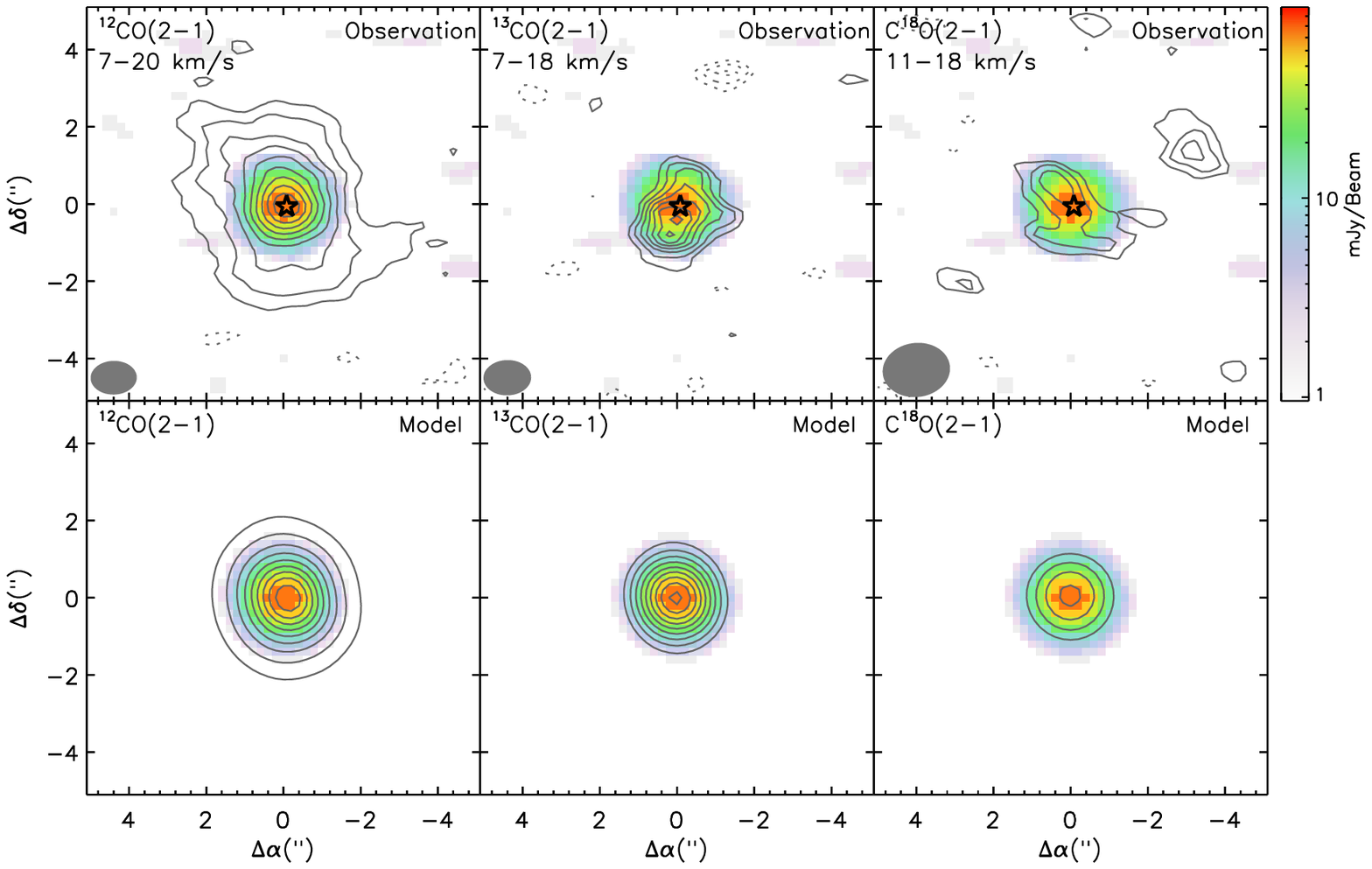}
\caption{Top panels: The velocity-integrated intensities (contours) for the $^{12}{\rm CO}~J=2-1$, $^{13}{\rm CO}~J=2-1$, and ${\rm C^{18}O}~J=2-1$ line emission, overlaid on the the continuum emission. For $^{12}{\rm CO}~J=2-1$,  the intensities are integrated over the velocity range between 6.5 and 20.8\,\kms, and the contours are drawn at 0.69\,Jy~beam$^{-1}$~\kms (3\,$\sigma$) intervals, starting at 0.69\,Jy~beam$^{-1}$~\kms  (3\,$\sigma$). For  $^{13}{\rm CO}~J=2-1$, the intensities are integrated over the velocity range between 7 and 20\,\kms, and the contours start at 0.48\,Jy~beam$^{-1}$~\kms (3\,$\sigma$) with an interval of  0.16\,Jy~beam$^{-1}$~\kms (1\,$\sigma$). For ${\rm C^{18}O}~J=2-1$, the intensities are integrated over the velocity range between 7 and 20\,\kms, and the contours begin at 0.24\,Jy~beam$^{-1}$~\kms (3\,$\sigma$),  and increase in  0.08\,Jy~beam$^{-1}$~\kms (1\,$\sigma$) increments. In each panel,  the negative contours, shown with the dashed lines, are drawn at  $-$1\,$\sigma$ intervals, starting at  $-$3\,$\sigma$. The synthesized beam for each line emission is shown in the lower left corner in each panel. Bottom panels: The modeled velocity-integrated intensities (contours) for the $^{12}{\rm CO}~J=2-1$, $^{13}{\rm CO}~J=2-1$, and ${\rm C^{18}O}~J=2-1$ line emission, overlaid on the modeled continuum emission, for the GW~Ori disk system. The contour levels are same as in the top panels at the corresponding molecular lines. \label{Fig:CO_cont}}
\end{figure*}

\subsection{Molecular line emission}\label{Sect:Molecular_line_emission}
In the top three panels in Fig.~\ref{Fig:CO_cont} we show the integrated intensity maps of $^{12}{\rm CO}~J=2-1$, $^{13}{\rm CO}~J=2-1$, and ${\rm C^{18}O}~J=2-1$ lines. The $^{12}{\rm CO}~J=2-1$ line emission map shows an elongated structure extended from the north-east to the south-west direction with a single peak coincided with the center of the continuum emission. The $^{12}{\rm CO}~J=2-1$ line emission map is clearly spatially resolved. A two-dimensional Gaussian fit to the  $^{12}{\rm CO}~J=2-1$ integrated intensity map gives an FWHM size of  2$\farcs$5$\times$3$\farcs$4 after the deconvolution of the synthesized beam, corresponding to 890$\times$1300\,AU at a distance 400\,pc,  which is much more extended than the continuum emission map. The  $^{13}{\rm CO}~J=2-1$ line integrated intensity map of GW~Ori is more compact than the  $^{12}{\rm CO}~J=2-1$ one, and comparable to the continuum emission map. From the disk of GW~Ori, we only marginally detect ${\rm C^{18}O}~J=2-1$ line.

\begin{figure*}
\centering
\includegraphics[width=1.\columnwidth]{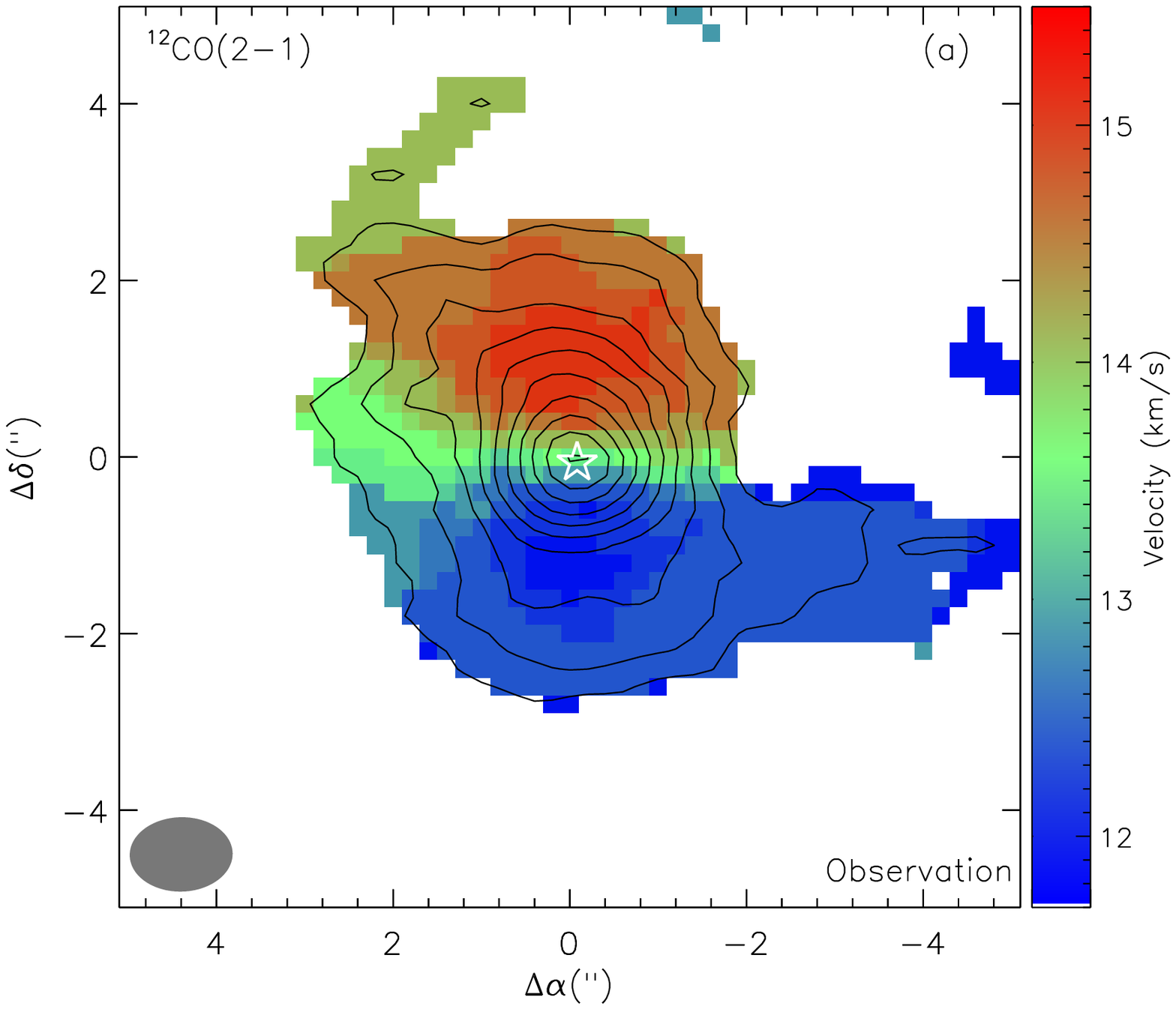}
\includegraphics[width=1.\columnwidth]{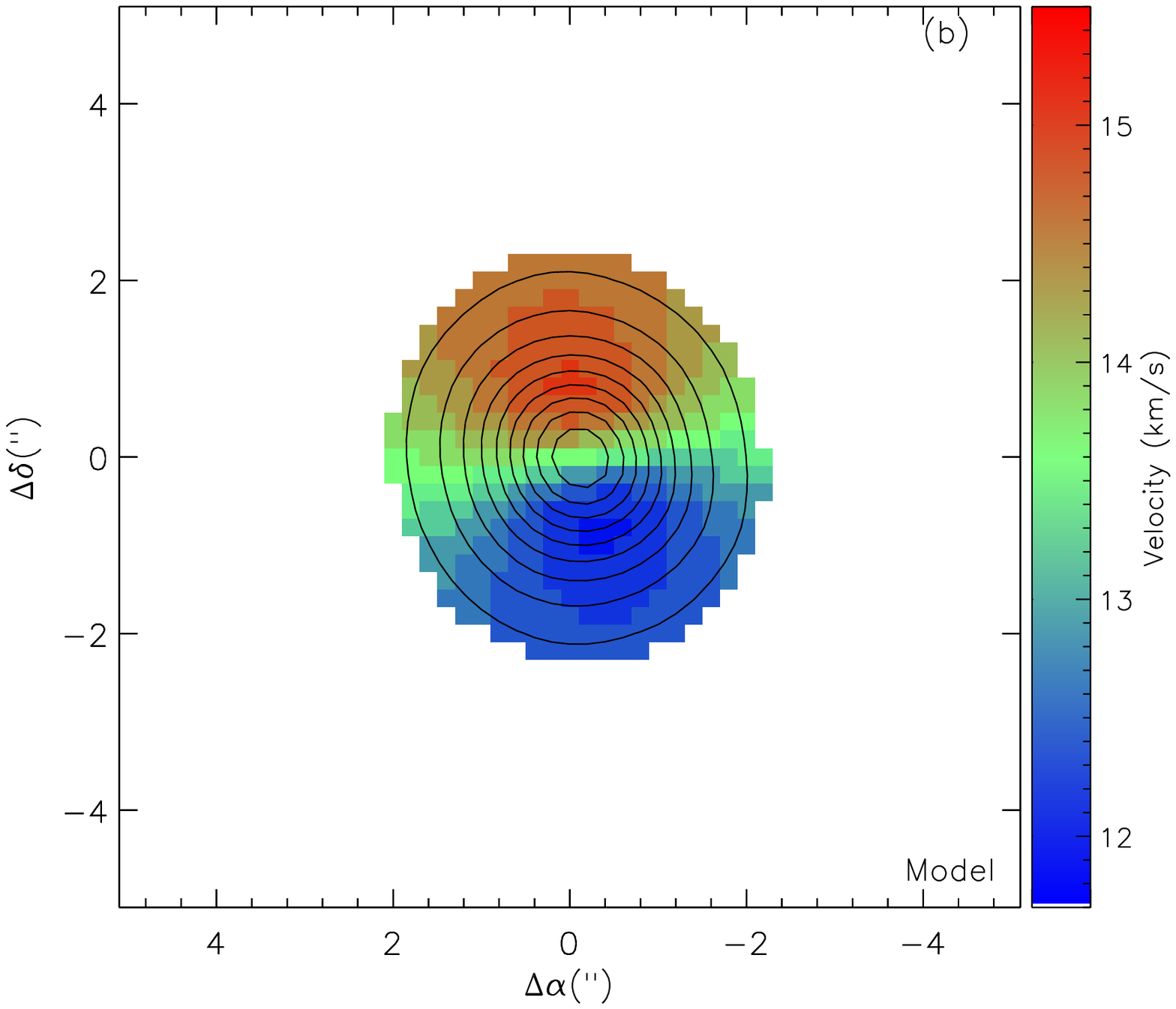}
\caption{(a) $^{12}{\rm CO}~J=2-1$ velocity (1st) moment map. The contours are for the  $^{12}{\rm CO}~J=2-1$ velocity-integrated intensities (see Fig.~\ref{Fig:CO_cont}), starting at 0.69\,Jy~beam$^{-1}$~\kms\ (3\,$\sigma$)  with an interval of 0.69\,Jy~beam$^{-1}$~\kms\ (3\,$\sigma$). The synthesized beam is shown in the lower left corner.  (b) Same as in Panel~(a), but for the predicted map from modeling.  \label{Fig:vel1}}
\end{figure*}

\begin{figure*}
\centering
\includegraphics[width=1.\columnwidth]{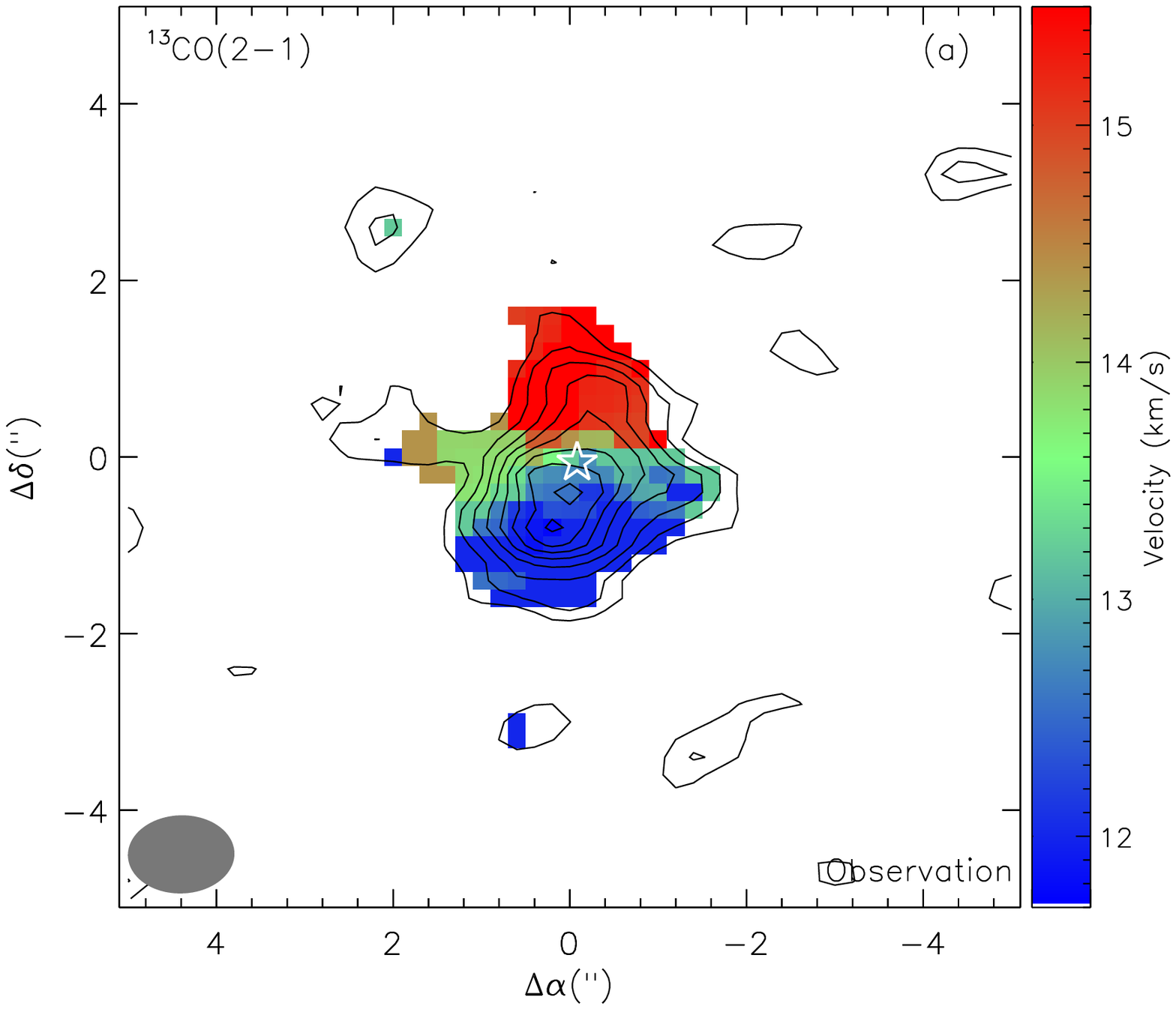}
\includegraphics[width=1.\columnwidth]{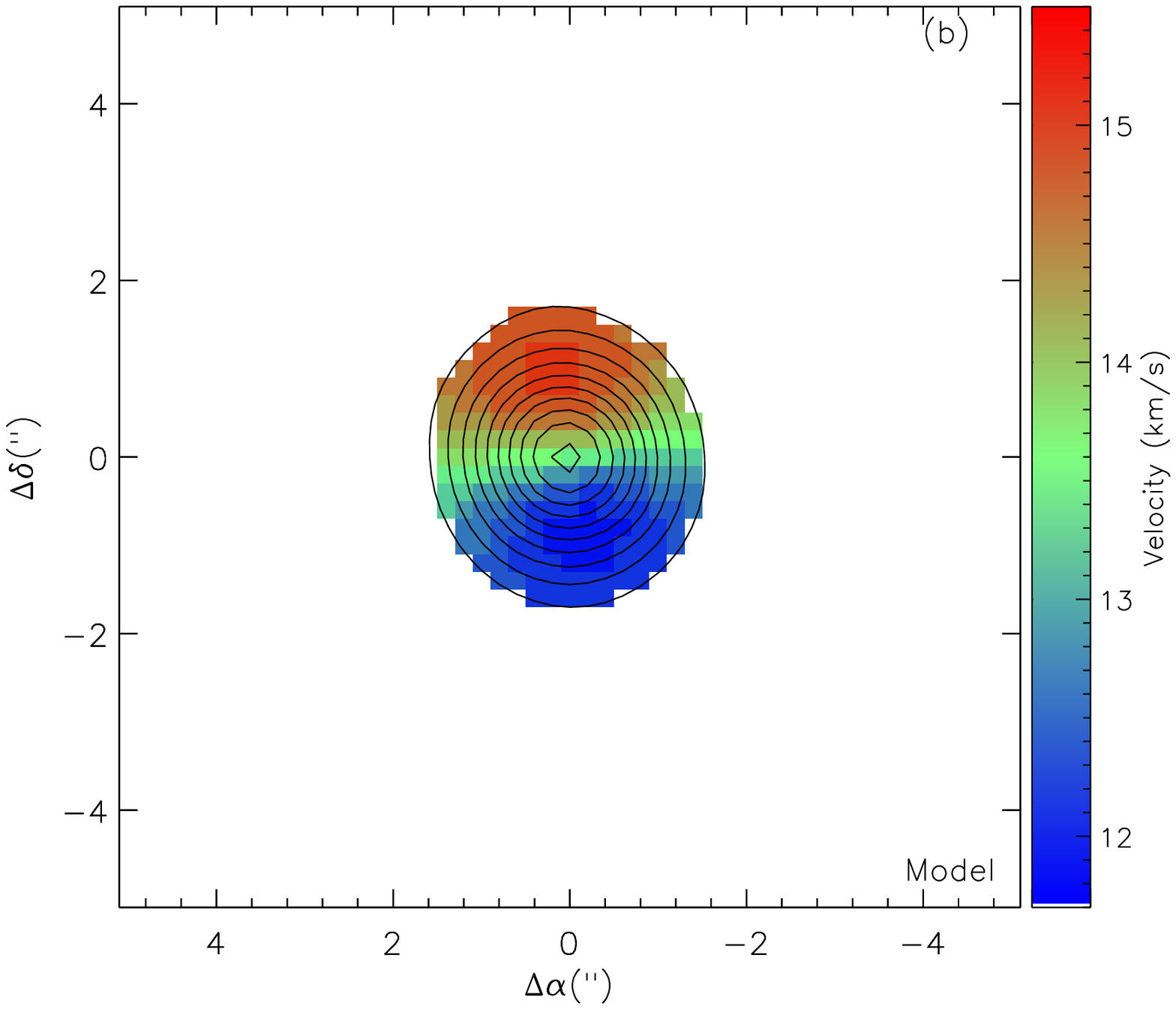}
\caption{(a) $^{13}{\rm CO}~J=2-1$ velocity (1st) moment maps. The contours are for the  $^{13}{\rm CO}~J=2-1$ velocity-integrated intensities (see Fig.~\ref{Fig:CO_cont}), starting at 0.48\,Jy~beam$^{-1}$~\kms\ (3\,$\sigma$)  with an interval of 0.16\,Jy~beam$^{-1}$~\kms\ (1\,$\sigma$).  The synthesized beam is shown in the lower left corner. (b) Same as in Panel~(a), but for the predicted map from modeling. \label{Fig:vel2}}
\end{figure*}

\begin{figure*}
\centering
\includegraphics[width=2.\columnwidth]{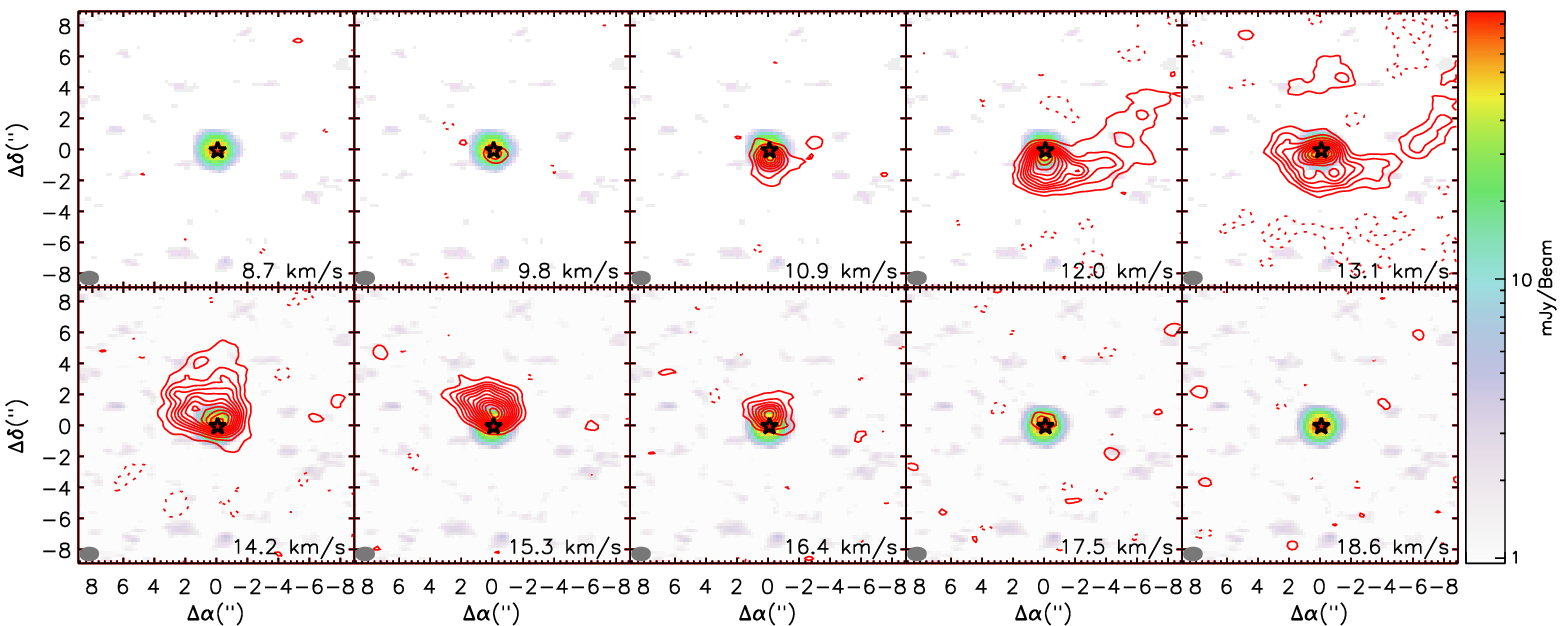}
\caption{The channel maps of the  $^{12}{\rm CO}~J=2-1$ emission toward GW~Ori. Channels are 1.1\,\kms\ wide with the synthesized beam marked in the bottom left corner. Contour levels are drawn at intervals of 0.127\,Jy~beam$^{-1}$ (3$\sigma$), starting at 0.127\,Jy~beam$^{-1}$. The dashed contours are the negative features with the same contours as the positive ones in each panel. The synthesized beam is shown in the lower left corner in each panel. \label{Fig:channel}}
\end{figure*}

In Fig.~\ref{Fig:vel1}(a) and Fig.~\ref{Fig:vel2}(a), we show the  $^{12}{\rm CO}~J=2-1$ and $^{13}{\rm CO}~J=2-1$ velocity (1st) moment maps, respectively, suggesting northern and southern parts of the disk are redshifted and blueshifted with respect to the relative velocity of GW~Ori, respectively. Fig.~\ref{Fig:channel} displays the channel maps of  $^{12}{\rm CO}~J=2-1$ line emission with contours starting at 0.127\,Jy~beam$^{-1}$ (3$\sigma$) with intervals of 0.127\,Jy~beam$^{-1}$. The velocity moment maps and the channel maps are generally consistent with the expected kinematic pattern for gas material in Keplerian rotation with substantial inclination to our line of sight. The disk inclination can be constrained using the position-velocity (PV) diagram of  the molecular lines. In Fig.~\ref{Fig:PV}(a), we present the PV diagram  from the $^{12}{\rm CO}~J=2-1$ map along the north-south direction cross the peak of the integrated intensity maps of $^{12}{\rm CO}~J=2-1$. In the figure, we plot the expected Keplerian rotation curves for a disk  inclined by 20$^{\circ}$, 40$^{\circ}$, and 60$^{\circ}$ around a star with a mass of 3.9\,\Msun\ for comparison. From the comparison, we infer that the disk inclination should be between 20$^{\circ}$ and 60$^{\circ}$.

Fig.~\ref{Fig:channel} shows the channel maps of  $^{12}{\rm CO}~J=2-1$ line emission. In the figure, at the channels with velocities of 12.0 and 13.1\,\kms, we note a tail structure originating from the outer disk and pointing to the north-western direction. Such a tail is also evident as the blueshifted structure in the  $^{12}{\rm CO}~J=2-1$  velocity moment map.  One explanation for the structure could be the cloud contamination which can be severe for CO lines. If it is the case, it is required that the velocity of the parental cloud around GW~Ori is $\sim$1\,\kms\ bluer than GW~Ori. A Gaussian fit to the $^{12}{\rm CO}~J=2-1$ spectrum at the peak of  {\newrev the integrated line intensity map} suggests that  the velocity of GW~Ori with respect to the local standard of rest (LSR) is around 13.6\,\kms, and the LSR velocity of the parental cloud of GW~Ori is around 12.7\,\kms\ \citep{2000A&A...357.1001L}, which supports the above explanation of the blueshifted tail structure.

\section{Disk modeling}\label{Sec:modeling}

{\newrev In this section, we use a simple disk model to reproduce the SMA observations of GW~Ori. Given that our data has low spatial resolution, any model involves a strong simplification of the very complex GW~Ori disk system \citep{2014A&A...570A.118F}. Furthermore, the complexity of disks means that the parameter space is highly degenerated and non-continuous \citep{2016PASA...33...59S}. Our aim is thus to explore the global gas and dust content of the GW~Ori disk in the light of typical disk models, and to compare it to other similar objects. Any further modeling is beyond the scope of this paper and worth only when higher resolution data becomes available. }

\subsection{Continuum emission}

\subsubsection{Parameters for modeling continuum emission}

 We define a global dust surface density in the same form as in \citet{2009ApJ...700.1502A},

\begin{equation}
\Sigma=\Sigma_{\rm c}~\bigg(\frac{R}{R_{\rm c}}\bigg)^{-\gamma}~{\rm exp}\Bigg[-\bigg(\frac{R}{R_{\rm c}}\bigg)^{2-\gamma}\Bigg],\label{Equ:sigma}
\end{equation}

\noindent where $\Sigma_{\rm c}$ is the normalization parameter at  the characteristic scaling radius $R_{\rm c}$, and $\gamma$  is the gradient parameter. The above profile, which has been used to successfully  model different types of disks in the literature \citep{2009ApJ...700.1502A,2011ApJ...732...42A,2012ApJ...744..162A}, is the similarity solution for a simple accretion disk with time-independent viscosity ($\nu$) and $\nu\propto R^{\gamma}$ \citep{1974MNRAS.168..603L,1998ApJ...495..385H}. In this work, we do not take $\Sigma_{\rm c}$ as the free parameter. Instead, we use the disk dust mass ($M_{\rm dust}$), and $\Sigma_{\rm c}$ can be calculated when we set up other parameters about the disk structure. We  set $\gamma$ as a free parameter. 

{\rev We include a vertical gradient in the dust size distribution in disk modelling, to simulate the dust settling in disks. In practice, we use two dust populations: small dust population, and large dust population, as did  \citet{2011ApJ...732...42A}. The dust density structure for each dust population in a spherical coordinate system ($R$, $\theta$, and $\phi$) is parameterized as 
 
\begin{equation}
\rho_{\rm small}=\frac{(1-f)\Sigma}{\sqrt{2\pi}Rh}~{\rm exp}\Bigg[-\frac{1}{2}\bigg(\frac{\pi/2-\theta}{h}\bigg)^{2} \Bigg],\label{Equ:rhos}
\end{equation}
\begin{equation}
\rho_{\rm large}=\frac{f\Sigma}{\sqrt{2\pi}R\Lambda h}~{\rm exp}\Bigg[-\frac{1}{2}\bigg(\frac{\pi/2-\theta}{\Lambda h}\bigg)^{2} \Bigg],\label{Equ:rhol}
\end{equation}

\noindent where $\rho_{\rm small}$ is the density for the small dust population, $\rho_{\rm large}$ is the one for the large dust population, and  $h$ is the  angular scale height. Following \citet{2011ApJ...732...42A}, we assume the large grains are distributed to 20\% of the scale height ($\Lambda$=0.2), and account for 85\% of the total column ($f$=0.85). We do not explore the parameter space of $\Lambda$ and $f$ since our observational data cannot provide an efficient constraint on them. The angular scale height $h$ is defined as 

\begin{equation}
 h=h_{\rm c}~\bigg(\frac{R}{R_{\rm c}}\bigg)^{\Psi},
\end{equation}

\noindent where $h_{\rm c}$ is the angular scale height at the  scaling radius $R_{\rm c}$, and $\Psi$ characterizes the flaring angle of the disk.

}

In Paper\,I, we have shown that a gap sized at 25--55\,AU needs to be included in the disk model. A small population of tiny dust particles (sizes 0.005--1\,\mum)  is also needed to distribute in the gap, in order to reproduce the moderate excess emission at near-infrared bands and the strong and sharp silicate feature at 10\,\mum\ on the SED of GW~Ori. In this work, we include a dust depletion factor ($\xi_{\rm gap}$) to modify the surface density within the gap. We simply set the gap size $R_{\rm gap}$=45\,AU, according to the results in Paper\,I, since our SMA data cannot provide any constraints on the inner disk. {\rev The inner radius of the gap ($R_{\rm in}$) is fixed to be 1.2\,AU, which is the orbital semi-major axis of the close companion GW~Ori~B (see Paper\,I).}  When the disk radius $R\le R_{\rm gap}$, the modified surface density is $\Sigma=\xi_{\rm gap}\Sigma$, where $\Sigma$ is obtained from Equation~\ref{Equ:sigma}.

\begin{figure*}
\centering
\includegraphics[width=1.\columnwidth]{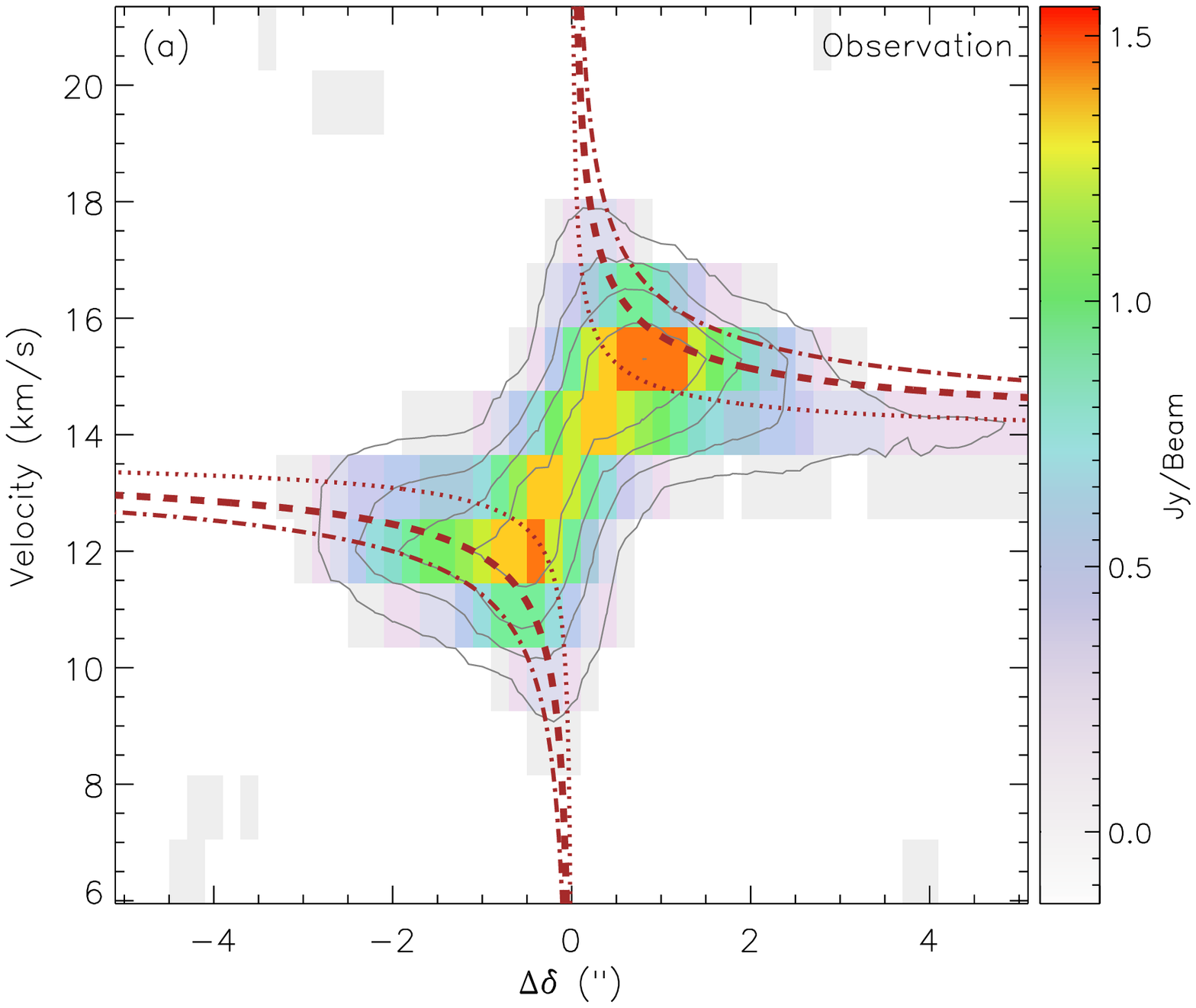}
\includegraphics[width=1.\columnwidth]{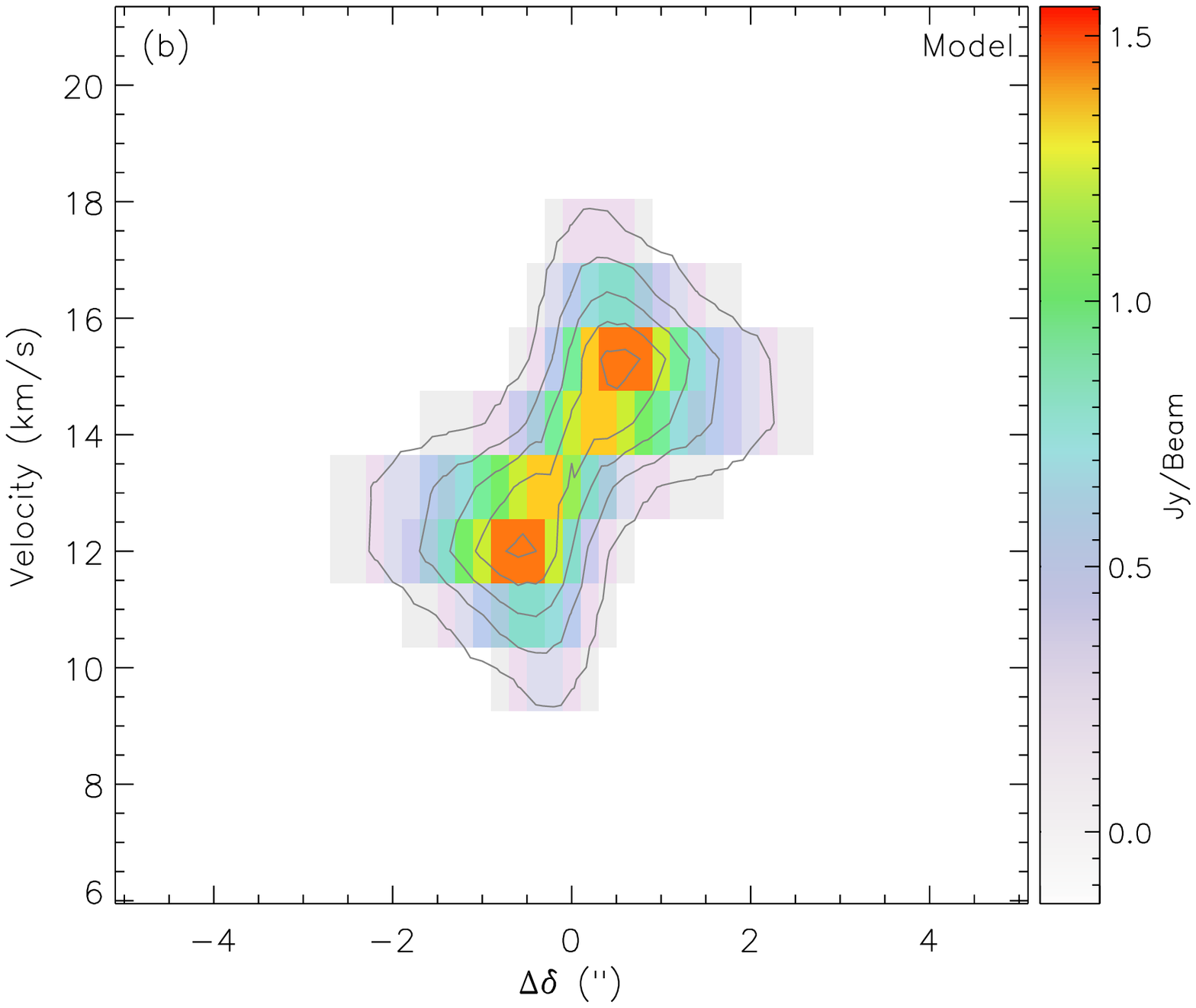}
\caption{(a) The observed position-velocity diagram  from  the $^{12}{\rm CO}~J=2-1$ map along the north-south direction cross GW~Ori. The offset refers to the distance along cut from GW~Ori. The contours start at 0.15\,Jy~beam$^{-1}$ (3$\sigma$) with an interval of 0.05\,Jy~beam$^{-1}$ (1$\sigma$). The dotted, dashed, and dash-dotted lines show a Keplerian rotation curve with a central mass of 3.9\,\Msun\ with disk inclinations of 20$^\circ$, 40$^\circ$, and 60$^\circ$, respectively. (b) Same as in Panel~(a) but for the modeled position-velocity diagram with a disk inclination of 35$^{\circ}$. \label{Fig:PV}}
\end{figure*}

In the disk models,  we use two populations of amorphous dust grains (25\% carbon and 75\% silicate) with a power-law size distribution with an exponent of $-3.5$ and  a minimum size of 0.005\,\mum. The maximum size of dust grains is set to be 1\,\mum\ in the gap, as suggested in Paper\,I, and {\rev to be 1\,\mum\ for the small dust population and 1000\,\mum\ for the large dust population in the outer disk ($R>R_{\rm gap}$). In Fig.~\ref{Fig:dust_opa}, we show the opacity spectra for the two dust populations derived from Mie calculations.} 

The stellar parameters adopted in the models are $T_{\rm eff}$=5500\,K, $R_{\star}$=7.6\,$R_{\odot}$, and $M_{\star}$= 3.9\,\Msun, taken from Paper\,I. We use the RADMC--3D code \citep[version~{\rev 0.40},][]{2012ascl.soft02015D} to do the radiative transfer in the disk models, and vary the free parameters to calculate the continuum emission maps at 1.3\,mm and  the SEDs. For simplicty, we convolve the model continuum emission maps with the synthetical beam for the SMA continuum emission map to simulate the observations.

{\rev \subsubsection{Scheme for modeling continuum emission}

In order to model the continuum emission from the disk of GW~Ori, we have 10 parameters. Among them, the 8 parameters,  $R_{\rm in}$, $R_{\rm gap}$, $\xi_{\rm gap}$, $R_{\rm c}$, $\gamma$, $h_{\rm c}$, $\Psi$, and $M_{\rm dust}$, are for the disk structure. As discussed above, we have fixed $R_{\rm in}$=1.2\,AU, $R_{\rm gap}$=45\,AU. To compare the model results with the observations, we need to know the orientation of the disk with respect to the observer, which can be characterized with two more parameters, the inclination ($i$) of the disk with respect to the line of sight ($i=90^\circ$ for edge-on disks), and the position angle (PA) of the disk major axis. The disk of GW~Ori is well resolved in the $^{12}{\rm CO}~J=2-1$ line. From the integrated intensity map of the $^{12}{\rm CO}~J=2-1$ line, we derive an inclination of $\sim$40$^{\circ}$ and a position angle$\sim$10$^{\circ}$. In disk modelling, we allow the inclination to change from 30$^{\circ}$, 40$^{\circ}$, to 50$^{\circ}$, and fix position angle to be 10$^{\circ}$.

The exploration of the parameter spaces is divided into three steps. At the first step, we calculate the continuum emission map at 1.3\,mm for a coarse and wide grid of parameters listed in Table~\ref{Tab:disk_par_grid}. Here we fix $\xi_{\rm gap}$=3$\times10^{-3}$, according to the SED modelling in Paper\,I, since it only insignificantly affects the 1.3\,mm continuum emission modelling. In total, we obtain 9360 modeled 1.3\,mm continuum emission maps. We compare the calculated continuum emission maps with the observation, which can efficiently constrain $M_{\rm dust}$ and $R_{\rm c}$. For each model, the goodness of the fit $\chi^2_{\rm mm}$ is calculated by

\begin{equation}
 \chi^2_{\rm mm}=\frac{\frac{1}{N}\sum^{N}_{i=1}(\mu_{i}-\omega_{i})^2}{\sigma^2} \label{Equ:equa5} 
\end{equation}

\noindent where $\sigma$ is the noise of the observed 1.3\,mm map, $N$ the number of pixels with the values above 3$\sigma$, $\mu$ the modeled data, and $\omega$ the observed data. By comparing with the observations, we obtain sets of disk models providing a good fit which we define as $\chi^2_{\rm mm}-\chi^2_{\rm mm,~best}<2$ where $\chi^2_{\rm mm,~best}$ is the minimum $\chi^2_{\rm mm}$ among all the models. In total, we have 102 good-fit models. Then, we calculate the SEDs of the good-fit disk models in the range from 70\,\mum\ to 1.3\,mm, and compare them with the one of GW~Ori. The goodness of the fit of the SED  $\chi^2_{\rm SED}$ is calculated by 
 
\begin{equation}
 \chi^2_{\rm SED}=\frac{1}{N}\sum^{N}_{i=1}\frac{(\mu_{i}-\omega_{i})^2}{\sigma_{i}^2}   
\end{equation}

\noindent where  $N$ is the number of the observed wavelengths, $\mu$ the synthetic flux density, $\omega$ the observed flux density, and  $\sigma$ the observational uncertainities. The total goodness of the fit $\chi^2_{\rm total}$ is given by 

\begin{equation}
 \chi^2_{\rm total}=\chi^2_{\rm mm}+\chi^2_{\rm SED}
\end{equation}

We derive the ranges for the free parameters of the models providing a good fit with  $\chi^2_{\rm total}-\chi^2_{\rm total,~best}<2$ where $\chi^2_{\rm total,~best}$ is the minimum $\chi^2_{\rm total}$ among the 102 models. The parameters of the good-fit models are listed in Table~\ref{Tab:disk_par_grid}.

\begin{figure}
\centering
\includegraphics[width=1.\columnwidth]{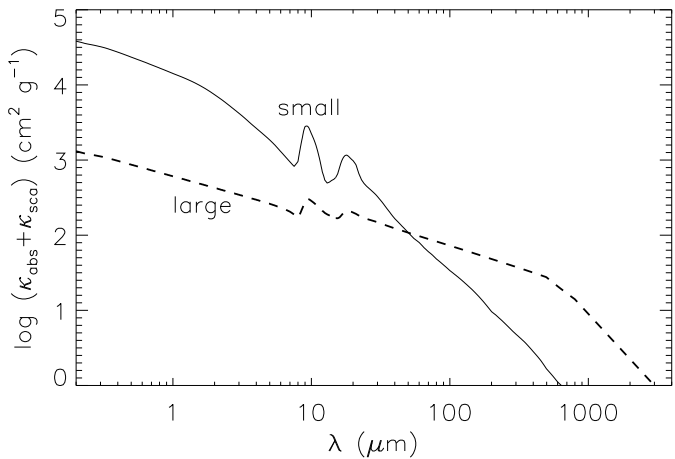}
\caption{Opacity spectra ($\kappa_{\rm abs}$ for absorption and $\kappa_{\rm sca}$ for scattering) for the small dust population (solid line) and the large dust population (dash line) used in our disk modeling. \label{Fig:dust_opa}}
\end{figure}

\begin{table*}
\caption{Coarse grids of the parameters for modeling disk continuum emission.\label{Tab:disk_par_grid}}
\centering
\begin{tabular}{ccc}
\hline\hline
            &        & Range  \\
Parameters  & Input  & ($\chi^2_{\rm total}-\chi^2_{\rm total,~best}<2$) \\
\hline
\multicolumn{3}{c}{{\bf Central star}}\\
\hline
Effective temperature ($T_{\rm eff}$)\tablefootmark{a} &\multicolumn{1}{c}{5500\,K}\\
Stellar radius\tablefootmark{a}  &\multicolumn{1}{c}{7.5\,$R_{\odot}$}\\
Stellar mass ($M_{\star}$)\tablefootmark{a}   &\multicolumn{1}{c}{3.9\,$M_{\odot}$}\\
\hline
\multicolumn{3}{c}{{\bf Disk}}\\
\hline
Inner radius ($R_{\rm in}$)\tablefootmark{a}             &1.2\,AU        \\ 
Gap size ($R_{\rm gap}$)\tablefootmark{a}                &45\,AU\\
Dust depletion factor ($\xi_{\rm gap}$)\tablefootmark{a}  &3$\times10^{-3}$     \\
Characteristic scaling radius ($R_{\rm c}$) &250, 270, 290, 310, 330, 350, 370, 390 \,AU  & 290--350\,AU \\
Surface density gradient parameter ($\gamma$) & 0.1, 0.3, 0.5, 0.7, 0.9, 1.1 & 0.1--0.5\\
Angular scale height ($h_{\rm c}$) & 0.15,0.17,0.19,0.21,0.23 & 0.15--0.19 \\
 $\Psi$\tablefootmark   &  0.1, 0.15, 0.2  &0.1--0.2 \\
Disk mass           &  [1, 1.5, 2, 2.5, 3, 3.5, 4, 4.5, 5, 5.5, 6, 6.7, 7]$\times10^{-2}\,M_{\star}$ &3--3.5$\times10^{-2}\,M_{\star}$ \\
Disk  inclination ($i$)  &    30$^\circ$, 40$^\circ$, 50$^\circ$  &  30$^\circ$--40$^\circ$\\
Position angle\tablefootmark{a}      &  10$^\circ$    \\
\hline
\end{tabular}
\tablefoot{\tablefoottext{a}{Fixed parameters.}}
\end{table*}

At the second step, we refine the grid of parameters with knowledge of the above good-fit models, and list them in Table~\ref{Tab:disk_par_refined_grid}. We repeat the procedure in the first step, but require $\chi^2_{\rm mm}-\chi^2_{\rm mm,~best}<0.5$ and  $\chi^2_{\rm total}-\chi^2_{\rm total,~best}<0.5$. We obtain  the ranges for the free parameters of the models providing a fit with  $\chi^2_{\rm total}-\chi^2_{\rm total,~best}<0.5$.  The disk model with minimum $\chi^2_{\rm total}$ is considered as the best-fit model. 

At the final step, we will constrain the dust depletion factor ($\xi_{\rm gap}$) by fitting the full SED of GW~Ori. We calculate a set of model SEDs by varing $\xi_{\rm gap}$ from 1$\times10^{-3}$ to 1$\times10^{-2}$ with other parameters from the above best-fit model. Since the change of $\xi_{\rm gap}$ can only obviously vary the shape of the SED within near- and mid-infrared wavelengths, we compare the model SEDs with the observed one within the within the wavelengths ranging from 1\,\mum\ to  37\,\mum. We find the model SEDs with $\xi_{\rm gap}=1.5-4\times10^{-3}$ can reproduce the observations, and the one with $\xi_{\rm gap}=2.5\times10^{-3}$ can give the best fit to the data.  {\newrev As mentioned before, our model is highly simplified and the low resolution does not allow us to expore in depth the degenerated and non-continuous parameter space of this very complex system. Thus the $\chi^2$ approach has to be taken as a tool to measure the goodness of fit, but it cannot be used to derive statistics on the significance of any particular model.}

\begin{table*}
\caption{Refined Grids of the parameters for modeling disk continuum emission.\label{Tab:disk_par_refined_grid}}
\centering
\begin{tabular}{cccc}
\hline\hline
            &    & Range  & \\
Parameters  & Input   & ($\chi^2_{\rm total}-\chi^2_{\rm total,~best}<0.5$) & Best-fit \\
\hline
Characteristic scaling radius ($R_{\rm c}$) &290, 300, 310, 320, 330, 340, 350 \,AU  &300--340\,AU  &320\,AU \\
Surface density gradient parameter ($\gamma$) & 0.1, 0.2, 0.3, 0.4, 0.5 &0.1--0.4  & 0.2\\
Angular scale height ($h_{\rm c}$) & 0.15, 0.16, 0.17, 0.18, 0.19  &0.15--0.19 & 0.18\\
 $\Psi$\tablefootmark   &  0.1, 0.15, 0.2 &0.1-0.2  &0.1 \\
Disk mass           &  [3.0, 3.1, 3.2, 3.3, 3.4, 3.5]$\times10^{-2}\,M_{\star}$ &3.0--3.4$\times10^{-2}\,M_{\star}$&3.1$\times10^{-2}\,M_{\star}$ \\
Disk  inclination ($i$)  &    30$^\circ$, 35$^\circ$, 40$^\circ$  & 30$^\circ$--35$^\circ$  & 35$^\circ$\\
\hline
\end{tabular}
\end{table*}

}

\subsection{CO emission}

{\rev With the parameters of the above best-fit model, we can compute the model  map of the $^{12}{\rm CO}~J=2-1$ velocity-integrated intensities, and compare it with the observation. However, to calculate the CO emission, one needs to know the gas temperaure distribution in the disk. At the first step, we assume that the gas and dust has equal temperatures, and ignore that CO molecules are frozen out of the gas phase when $T_{\rm gas}<20$\,K and  that the gas temperatures may be higher than the dust temperatures near the disk surface. The velocity fields of gas material in the disk models are assumed to be Keplerian. The  thermal line broadening is automatically included in the RADMC--3D code. Besides it, we also include the turbulent line broadening by assuming a constant linewidth of 0.01\,\kms\ from turbulence. Taking {\newrev an abundance ratio}  $^{12}{\rm CO}/H_{2}=10^{-4}$, which is the canonical abundance of interstellar medium (ISM), we calculate the $^{12}{\rm CO}~J=2-1$ line emission using the RADMC–3D code with an assumption of local thermal equilibrium (LTE) conditions. The resulting maps are simply convolved with the corresponding synthetical beams, and then compared with the observation. We find that the predicted peak intensity of $^{12}{\rm CO}~J=2-1$  is about 3 time weaker than the observation. We vary the parameters among the good-fit models listed in Table~\ref{Tab:disk_par_refined_grid}, and the model emission for $^{12}{\rm CO}~J=2-1$ line are all  2--3  time weaker than the observation. The previous studies have shown that the gas temperatures could exceed the dust temperatures in the disk surface layers \citep{2006ApJ...636L.157Q,2009A&A...501..269P}, possibly due to additional ultraviolet or X-ray heating from central stars \citep{2004ApJ...615..972G,2004A&A...428..511J}. In order to reproduce the observed CO emissions of GW~Ori, following \citet{2012ApJ...744..162A},  we parameterize the gas temperature as}

\begin{equation}
T_{\rm gas} = \left\{ 
  \begin{array}{l l}
    T_{\rm atm}+(T_{\rm mid}-T_{\rm atm})~\Bigg[{\rm cos}\bigg(\frac{\pi}{2}\frac{\pi/2-\theta}{h_{\rm q}}\bigg)\bigg]^{2\delta} & \quad \text{if $\pi/2-\theta<h_{q}$}\\
    T_{\rm atm} & \quad \text{if $\pi/2-\theta\ge h_{q}$}
  \end{array} \right.
\end{equation}

\noindent Where $T_{\rm mid}$ is the midplane temperature derived from the RADMC--3D simulations of the dust, $\delta$ describes the steepness of the vertical profile, and $h_{\rm q}=4H_{\rm p}$ where $H_{\rm p}$ is the angular pressure scale height determined from  $T_{\rm mid}$. We calculate $H_{\rm p}$ as $H_{\rm p}=(kT_{\rm mid}R/GM_{\star}\mu m_{\rm H})^{1/2}$ where $k$ is the Bolzmann's constant, $G$ is the gravitational constant, and $\mu=2.37$ is the mean molecular weight of the gas. $T_{\rm atm}$ is the temperature in the disk atmosphere, parameterized as
\begin{equation}
T_{\rm atm}=T_{\rm atm, 100\,AU}\bigg(\frac{R}{\rm 100\,AU}\bigg)^{\zeta}
\end{equation}

\noindent For gas temperatures, we only set $T_{\rm atm, 100\,AU}$ as a free parameter and fix $\delta=2$ and $\zeta$=$-$0.5. We assume that the CO molecules are frozen out of the gas phase when $T_{\rm gas}<20$\,K. {\rev Since $^{12}{\rm CO}~J=2-1$ line emission can be used to constrain the disk inclination better than dust continuum emission, we calculate the $^{12}{\rm CO}~J=2-1$ line emission using three disk inclinations, 30$^\circ$, 35$^\circ$, and 40$^\circ$. The $^{12}{\rm CO}~J=2-1$ line emission are computed using RADMC-3d assuming non-local thermal equilibrium (non-LTE) conditions for different $T_{\rm atm, 100\,AU}$. The result integrated intensity maps of  $^{12}{\rm CO}~J=2-1$ are simply convolved with the corresponding synthetical beams, and then compared with the observations to characterize  $T_{\rm atm, 100\,AU}$. For each model, the goodness of the fit $\chi^2_{\rm CO}$ is calculated in the similar way as Equation~\ref{Equ:equa5}, but only for the pixels with values above half of the peak intensity to reduce the possibility of CO contamination by the parental cloud. We also compare the PV diagrams from the observation and from the models to constrain the disk inclination. from a $\chi^2$ test, we find a disk model with $T_{\rm atm, 100\,AU}$=200$\pm$10\,K and $i=35^\circ-40^\circ$ can fit the observations. In Table~\ref{Tab:disk_par}, we list the best-fit parameter for modeling continuum and gas emission of GW~Ori.}

\subsection{Model results}\label{Sect:model_ressult}

In our disk models, we have {\rev 8} free parameters. The dust depletion factor ($\xi_{\rm gap}$) are mainly constrained by comparing the model SEDs with the observed one at near- and mid-infrared wavelengths, $h_{\rm c}$ can be estimated by fitting the SED at mid- and far-infrared bands, and $M_{\rm dust}$ can be constrained by fitting the SED of GW~Ori at submillimeter and millimeter wavelengths and the 1.3\,mm continnum emission map. The parameters $R_{\rm c}$ and  $\gamma$  are mainly constrained by comparing the model continuum emission map at 1.3\,mm with the observations. {\rev The disk inclination ($i$) is  constrained by fitting the  1.3\,mm  continuum emission map,  the  $^{12}{\rm CO}~J=2-1$ line emission map, and  the PV diagram. And the gas temperature parameter $T_{\rm atm, 100\,AU}$ is constrained by fitting the  $^{12}{\rm CO}~J=2-1$ line emission map. } In Table~\ref{Tab:disk_par}, we list the disk model which can satisfactorily reproduce both the 1.3\,mm continnum emission, the SED, and the $^{12}{\rm CO}~J=2-1$ line emission of GW~Ori. In the following, we compare the model results using these parameters with the observations.

\begin{table}
\caption{Disk model parameters for modeling dust and gas emission of GW~Ori.\label{Tab:disk_par}}
\centering
\begin{tabular}{cc}
\hline\hline
Parameters  & Values   \\
\hline
\multicolumn{2}{c}{{\bf Central star}}\\
\hline 
Effective temperature\tablefootmark{a} &\multicolumn{1}{c}{5500\,K}\\
Stellar radius\tablefootmark{a}  &\multicolumn{1}{c}{7.5\,$R_{\odot}$}\\
Stellar mass\tablefootmark{a}   &\multicolumn{1}{c}{3.9\,$M_{\odot}$}\\
\hline
\multicolumn{2}{c}{{\bf Disk}}\\
\hline
Inner radius ($R_{\rm in}$)\tablefootmark{a}            &1.2\,AU        \\ 
Gap size ($R_{\rm gap}$)\tablefootmark{a}                &45\,AU\\
Dust depletion factor ($\xi_{\rm gap}$)  &2.5$\times10^{-3}$     \\
Characteristic scaling radius ($R_{\rm c}$) &320\,AU  \\
Surface density gradient parameter ($\gamma$) & 0.2  \\
Angular scale height ($h_{\rm c}$) & 0.18  \\
 $\Psi$                    &  0.1   \\
Disk mass           &  0.12\,\Msun  \\
$T_{\rm atm, 100\,AU}$ &  200\,K\\  
Power index for $T_{\rm atm}$ ($\zeta$)\tablefootmark{a}   &$-$0.5   \\
$\delta$\tablefootmark{a}      & 2   \\
Disk  inclination ($i$)  &    35$^\circ$  \\
Position angle\tablefootmark{a}   &  10$^\circ$    \\
\hline
\end{tabular}
\tablefoot{\tablefoottext{a}{Fixed parameters.}}
\end{table}

\begin{figure}
\centering
\includegraphics[width=1.\columnwidth]{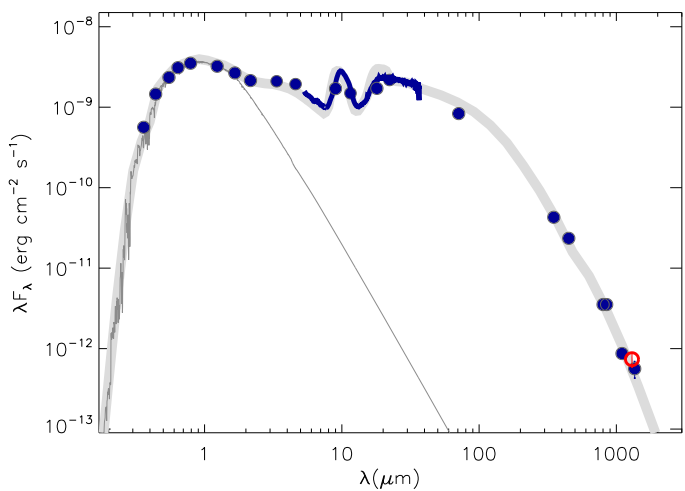}
\caption{The observed SED of GW~Ori. The broad band photometry is shown with the filled circles, and the IRS spectrum of this source is displayed in solid line. The open circle show the flux from our SMA observations. The thick gray line shows our model SED. The photospheric emission level is indicated with a thin gray curve.\label{Fig:SED}}
\end{figure}

In Fig.~\ref{Fig:cont}(b), we show the model continuum emission at 1.3\,mm. In Fig.~\ref{Fig:cont}(c, d), we compare the distribution of the intensity  along the east-western and  south-northern direction across the center of the map, respectively, from the model. The model results can fit the observations well. In Fig.~\ref{Fig:SED}, we compare the model SED with the observed one. The observed SED in the figure is the type~1 SED for GW~Ori in Paper\,I, and constructed using the $UBVR_{\rm C}I_{\rm C}$ photometry from \citet{2004AJ....128.1294C}, the $JHK_{\rm s}$ photometry from the 2MASS survey \citep{2006AJ....131.1163S}, the photometry at 3.4, 4.6, 12, and 22\,\mum\ from the WISE survey \citep{2010AJ....140.1868W}, the photometry at 9 and 18\,\mum\ from the AKARI survey \citep{2010A&A...514A...1I}, the MIPS 70\,\mum\ photometry from Paper\,I, the fluxes at 350, 450, 800, 850, 1100, 1360\,\mum\ from \citet{1995AJ....109.2655M}, and the 5$-$37\,\mum\  low-resolution IRS spectrum from paper\,I. In Fig.~\ref{Fig:SED}, it can be seen that our simple disk model can well reproduce the observed SED of GW~Ori. In Fig.~\ref{Fig:CO_cont}, we show the model map of the $^{12}{\rm CO}~J=2-1$ velocity-integrated intensities calculated with the parameters in  Table~\ref{Tab:disk_par}, which reproduces the observation very well. Figure~\ref{Fig:vel1}(b) displays the model $^{12}{\rm CO}~J=2-1$ velocity moment map, and Fig.~\ref{Fig:PV}(b) shows the model PV diagram for the  $^{12}{\rm CO}~J=2-1$ lines\footnote{The LSR velocity of GW~Ori is assumed to be 13.6\,\kms\ (see Sect.~\ref{Sect:Molecular_line_emission}).}.

Using the disk parameters listed in Table~\ref{Tab:disk_par}, and taking the typical abundances of ISM for $^{13}{\rm CO}$ and ${\rm C^{18}O}$, $^{13}{\rm CO}/{\rm H_{2}}=1.43\times10^{-6}$ and ${\rm C^{18}O}/{\rm H_{2}}=1.82\times10^{-7}$, i.e. $^{12}{\rm CO}/^{13}{\rm CO}$=70 and $^{12}{\rm CO}/{\rm C^{18}O}$=550 \citep{1994ARA&A..32..191W}, we calculate the the predicted  emission of the $^{13}{\rm CO}$ and ${\rm C^{18}O}$ lines  using RADMC--3D assuming Non-LTE conditions. {\rev We note the predicted line emission of $^{13}{\rm CO}~J=2-1$ is consistent with the observations (see  Fig.~\ref{Fig:CO_cont} and Figure~\ref{Fig:vel2}(b)), but the model line emission of ${\rm C^{18}O}~J=2-1$ is three time stronger than the observational data. A simple solution to reduce the model line emission is to decrease the abundance of ${\rm C^{18}O}$ in gas phase, which can be due to the real reduction in their abundances or more freezing than we assumed. We find that a satisfactory fit to the SMA data required that ${\rm C^{18}O}/{\rm H_{2}}\sim2.3\times10^{-8}$. In Fig.~\ref{Fig:CO_cont}, we show the model map of ${\rm C^{18}O}~J=2-1$  velocity-integrated intensities calculated using the above abundances. Such global ${\rm C^{18}O}$ gas-phase depletion in circumstellar disks has been suggested before \citep{1994A&A...286..149D,1996A&A...309..493D,2003A&A...399..773D,2007A&A...469..213I}, which can be due to a selective photodissociation of CO and its isotopes in disks \citep{2009A&A...503..323V}. However, we must stress that the constraint on the abundance of ${\rm C^{18}O}$ in the disk of GW~Ori based on the SMA data is just tentative since the detection of ${\rm C^{18}O}$ is marginal. Furthermore, an underestimate of the dust absorption of the line emission in the disk modelling or over-subtraction of continuum around  ${\rm C^{18}O}~J=2-1$ could complicate the issue.}

We have shown that our disk model can reproduce the SMA observations of GW~Ori. However, multi-parameter disk models are known to be highly degenerated and non-continuous. In addition, there are other sources of uncertain in disk modeling. The dust growth and settling in disks can change the dust properties vertically and radially \citep{2010A&A...513A..79B,2012A&A...539A.148B}. And it is unknown whether the gas and dust are well mixed in the GW~Ori disk, and the gas-to-dust ratio may vary vertically and radially \citep{2010A&A...513A..79B}. All these put strong limitations on the interpretations of the results from our disk modeling. The inferred disk masses can be strongly dependent on the assumed dust model (size distribution) and gas-to-dust ratio. {\rev In addition, it should be expected that the structure of the disk, vertical scale height, dust distribution, and heating may differ from typical cases due to the interaction between the two companions and the very massive disk in GW~Ori.} 


\section{Discussion}\label{Sec:discussion}

The disk inclination of GW~Ori is constrained by the gas kinematics in the disk traced by the $^{12}{\rm CO}$ line, which gives an intermediate inclination ($\sim$35$^\circ$). In Paper\,I we have estimated the inclination of the stellar rotation axis of the primary star (GW~Ori~A), which is around 35--50$^\circ$. Thus the stellar rotation axis of GW~Ori~A and the disk spin axis could be aligned. It is still unclear if the  the binary orbital plane and the disk is aligned in the GW~Ori system. If it is the case, the mass of the close companion GW~Ori~B is estimated to be {\rev 0.44}\,\Msun\ using the minimum companion mass ($m_{2}{\rm sin}~i_{*}=0.25$, $i_*$ is the inclination of the orbit) derived in Paper\,I. In this case, the expected $H$-band flux ratio between the primary GW~Ori~A (3.9\,\Msun) and GW~Ori~B (0.44\,\Msun) is 30:1 at age of $\sim$0.9\,Myr from the pre-main-sequence evolutionary tracks of \citet{2000A&A...358..593S}. On the contrary, the near-infrared interferometric observations show that the two stellar components may have near-equal $H$-band fluxes (2:1), which requires a low inclination of the orbit ($\sim$10$^{\circ}$) for  GW~Ori to have a massive close companion GW~Ori~B. {\rev However, an inclined orbit of a massive companion can drastically disturb the disk \citep{1996MNRAS.282..597L}.}  In addition, \citet{1998AstL...24..528S} detected the eclipses toward GW~Ori during 1987-1992, which may suggest a nearly edge-on inclination of the the orbit for  GW~Ori~A/B. {\rev However, recent observation with Kepler/K2 observations detect many quasi-periodic or aperiodic dimming events from young stars with disks, which are not edge-on, and could be due to inclined and variable inner dust disk warps \citep{2016ApJ...816...69A,2016MNRAS.462L.101A,2016MNRAS.463.2265S}. Thus, an intermediate disk inclination of GW~Ori does not contradict with the observation from \citet{1998AstL...24..528S}.}

\section{Summary}\label{Sec:summary}

Using the SMA we have mapped the disk around GW~Ori both in continuum and in the $J=2-1$ transitions of $^{12}{\rm CO}$, $^{13}{\rm CO}$, and ${\rm C^{18}O}$. The dust and gas properties in the disk are obtained by comparing the observations with the predictions from disk models with various parameters. 

We find a clear evidence that the circumstellar material in the disk is in Keplerian rotation around GW~Ori with a disk inclination of $\sim$35$^\circ$. 

{\rev We present a disk model which can  reproduce the dust continuum and line emission of CO and its isotopes from the disk of GW~Ori. To reproduce the line emission of ${\rm C^{18}O}$, we may need the substantially depleted abundances of ${\rm C^{18}O}$ in gas phase.}

GW~Ori is one of the most remarkable disks regarding its mass, and one of the most remarkable stellar systems (a massive G8 star with two companions). This object is bright at the whole electromagnetic spectrum and well studied, and an ideal target for future observations with ALMA.

\begin{acknowledgements} 
MF acknowledges support of the action “Proyectos de Investigación fundamental no orientada”, grant number AYA2012-35008. ASA support of the Spanish MICINN/MINECO “Ramón y Cajal” program, grant number RYC-2010-06164, and the action “Proyectos de Investigación fundamental no orientada”, grant number AYA2012-35008. YW acknowledges the support by NSFC through grants 11303097. This research has made use of the SIMBAD database, operated at CDS, Strasbourg, France. This publication makes use of data products from the Two Micron All Sky Survey, which is a joint project of the University of Massachusetts and the Infrared Processing and Analysis Center/California Institute of Technology, funded by the National Aeronautics and Space Administration and the National Science Foundation. This publication makes use of data products from the Wide-field Infrared Survey Explorer, which is a joint project of the University of California, Los Angeles, and the Jet Propulsion Laboratory/California Institute of Technology, funded by the National Aeronautics and Space Administration. This research is based on observations with AKARI, a JAXA project with the participation of ESA. This work is in part  based  on observations made with the Spitzer Space Telescope, which is operated by the Jet Propulsion Laboratory, California Institute of Technology under a contract with NASA. 
\end{acknowledgements}

\bibliographystyle{aa}
\bibliography{references}

\end{document}